\documentclass[11pt]{article}

\pdfoutput=1

\usepackage[left=1.5cm,right=1.5cm,top=20mm,bottom=3cm, columnsep=20pt]{geometry} 
\usepackage{blindtext}

\usepackage[english]{babel} 
\usepackage[utf8]{inputenc} 

\usepackage{amsmath}        
\usepackage{amsfonts}       
\usepackage{amssymb}

\usepackage{url} 	    
\usepackage{float}      
\usepackage{graphicx}   
\usepackage{subfigure}  
\usepackage[font={small}]{caption} 

\usepackage[square,super,compress]{natbib}      

\usepackage{hyperref}
\hypersetup{
    colorlinks=true,
    linkcolor=blue,
    filecolor=magenta,      
    urlcolor=blue,
    }

\date{}

\title  {\textbf{Analysis of the robustness and dynamics of spin-locking preparations for the detection of oscillatory magnetic fields}}
\author{
    Milena Capiglioni$^{1}$\footnote{\href{mailto:milena.capiglioni@extern.insel.ch}{milena.capiglioni@extern.insel.ch}} , Federico Turco$^{1}$, Roland Wiest$^{1}$, Claus Kiefer$^{1}$ \\
    \small{$^{1}$Institute for Diagnostic and Interventional Neuroradiology, Support Centre for Advanced Neuroimaging (SCAN),}\\\small{University of Bern, Bern, Switzerland.} \\
}

\begin{document}

\maketitle


\begin{abstract}
     Extracting quantitative information of neuronal signals by non-invasive imaging is an outstanding challenge for understanding brain function and pathology. However, state-of-the-art techniques offer low sensitivity to deep electrical sources. Stimulus induced rotary saturation (SIRS) is a recently proposed magnetic resonance imaging (MRI) sequence that detects oscillatory magnetic fields using a spin-lock preparation. Phantom experiments and simulations proved its efficiency and sensitivity, but the susceptibility of the method to field inhomogeneities is still not well understood. In this study, we simulated and analyzed the dynamic of three spin-lock preparations and their response to field inhomogeneities in the presence of a resonant oscillating field. We show that the composite spin-lock preparation is more robust against field variations within the double resonance effect. In addition, we tested the capability of the chosen composite spin-lock preparation to recover information about the spectral components of a composite signal. This study sets the bases to move one step further towards the clinical application of MR-based neuronal current imaging. 
\end{abstract}


\bigskip


\section{Introduction} \label{sec:intro}

Neuroscience seeks to understand brain activity and its pathologies but still struggles to extract electrical neuronal information in a non-invasive way. Non-invasive detection methods such as electroencephalogram \citep{Michel2012} and magnetoencephalogram \citep{mmg1993} measure electric potential and magnetic fields on the scalp and reconstruct the location of the electric sources. Unfortunately, the sensitivity of these methods is limited for in-depth activity sources \citep{Barkley2003,Ahlfors2010}. On the other hand, conventional functional magnetic resonance (fMRI) detects $T_2$ and  $T_2^*$ changes generated by blood oxygenation levels on the brain (BOLD contrast) \citep{Ogawa1990}. This technique presents good spatial resolution, but it measures neuronal activity indirectly, and the hemodynamic response function limits its temporal resolution \citep{Logothetis2001}.

Stimulus Induced Rotational Saturation (SIRS) is an MRI-based sequence with the potential to directly detect neuronal activity with sufficient spatial resolution \citep{Witzel2008}. The SIRS contrast is based on the interaction between the magnetization locked by a spin-lock (SL) pulse and an oscillating magnetic field induced by neuronal activity. Several studies proved the capabilities of the method with simulations and phantom experiments \citep{Witzel2008,Halpern2010,Sheng2016,Ueda2020,NAGAHARA2013}. Even more, the dependence of the contrast with the change of the excitation angle has been proposed as a new method for direct functional connectivity measurements \citep{Ito2020}.

The susceptibility of the SL preparations to $B_0$ and $B_1$ inhomogeneities hinders the further development and clinical application of this approach \citep{Ito2020,Sogabe2020}. However, researchers in the field of $T_1\rho$ relaxometry thoroughly studied the use of composite SL pulses to reduce the effect of field imperfections \citep{Charagundla2003,Witschey2007,Wang2015,Yuan2012}. Furthermore, some researchers within the SIRS community tested and used composite pulses \citep{Halpern2010,Sheng2016,Jiang2016}. But, to the best of our knowledge, no comprehensive study of the influence of the double resonance effect in combination with field inhomogeneities has been presented. Magnetization dynamics during a spin-lock pulse changes radically in the presence of an oscillating resonant field. Here, we present a simulation model based on rotation matrices that allow us to study the new dynamics and capture the effect of any oscillating field over time. We studied three well known SL preparations, the basic or standard SL pulse (BASL), the rotary echo SL pulse (RESL), and the composite rotary echo SL pulse (CRESL). Using the simulator and phantom experiments, we define the optimal strategy by examining the response of the preparations to field imperfections considering the new dynamics.

Finally, to validate the chosen preparation in more realistic conditions, we studied the filter properties of the preparation in the presence of composite signals. Physiological signals fluctuate in time and are composed of multiple frequencies \citep{Gonzales2000,Buzsaki2012,Wang2010}. In addition, certain pathologies manifest with neuronal activity at specific frequencies. For example, high-frequency oscillations (HFOs) are considered markers of the seizure onset zone in epilepsy \citep{Khosravani2009,Charupanit2020}. This case is noteworthy because non-invasively outlining the seizure onset zone can have a direct impact on patient care quality. Therefore, we consider that the next step to take towards the clinical application of the SIRS sequence is a more realistic signal analysis. The use of a repetitive acquisition of successive measurements with and without SL preparation has previously been proposed \citep{Kiefer2016}. With this method, the BOLD effect and image artifacts are minimized by dividing the two signals. In the second part of this work, we tested the ability of this acquisition method to reconstruct the frequency distribution of multi-frequency signals.

The purpose of this study is to advance research towards the clinical application of the SIRS technique. First, we define the optimal SL preparation and metric to detect oscillating fields in the presence of field inhomogeneities and second, we analyzed the response of the chosen preparation to composite signals.

\section{Methods}

\subsection{Magnetization dynamics under spin-lock preparations schemes} 

\begin{figure}[H]
    \centering
    \includegraphics[width=.9\columnwidth]{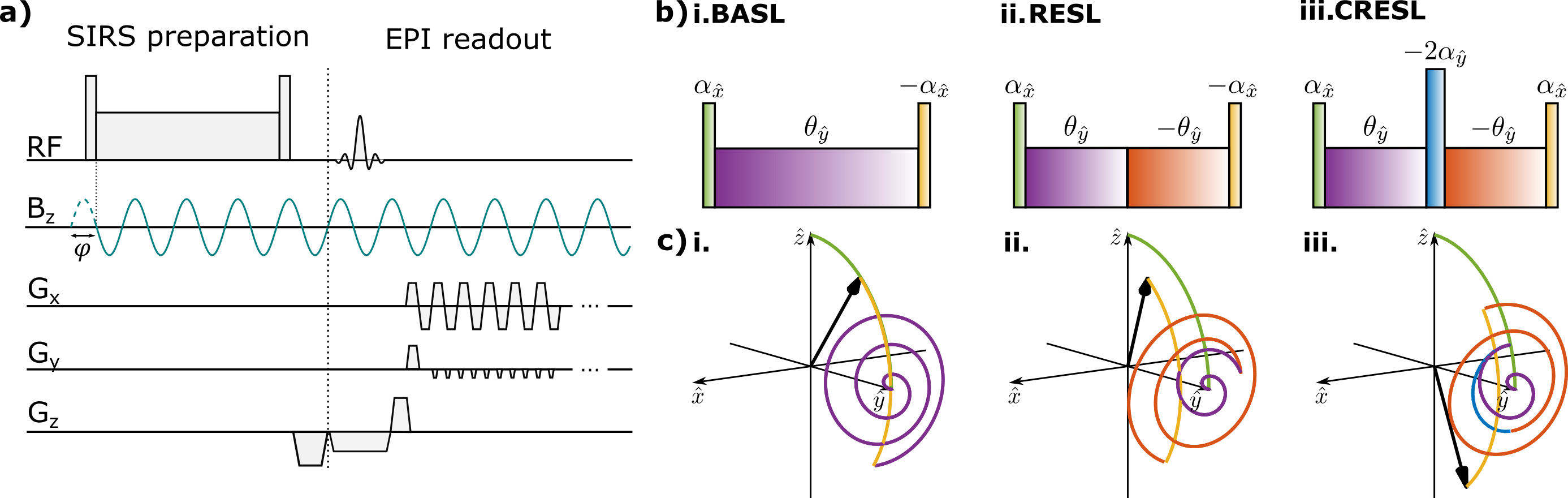} 
    \caption{Pulse sequence, SL preparations and magnetization dynamics. a) Pulse sequence diagram for a conventional SIRS preparation (BASL) followed by an echo planar imaging (EPI) acquisition. An exemplary oscillating target field ($B_z$) is shown, the initial phase ($\varphi$) corresponds to the phase at the beginning of the SL pulse. b) RF pulses that define the three SL preparations studied in this work: i. basic SL preparation (BASL), ii. rotary echo SL pulse (RESL) and iii. composite rotary echo SL pulse (CRESL). c) Representative diagram of the magnetization dynamic for the SL preparations showed in b) when in the presence of a resonant oscillating field applied along $\hat{z}$. The trajectory is color-coded to differentiate the effect of each RF pulse}
    \label{fig:1_Prep_dynamics}
\end{figure}

The dynamics of the magnetization with the BASL preparation in the presence of simple sinusoidal oscillating fields has been analytically described \citep{Ueda2018}. However, modulated oscillating fields are more representative of those found in physiological activity. In general, it is not possible to analytically describe these fields, and numerical simulations are the only option. Among the numerical simulation methods, solving differential equations \citep{Sheng2016} and step-wise constant rotations as presented in this work, give similar results. Based on previous bibliography in the $T_{1\rho}$ field \citep{Witschey2007,Wang2015,Yuan2012}, where composite and refocusing SL pulses were developed, we decided to adopt the rotation matrix method presented in those works and added the influence of an arbitrary oscillating target field. Figure \ref{fig:1_Prep_dynamics} (a) shows the conventional SIRS sequence formed by a spin-locking pulse followed by a single shot EPI readout. A traditional SL pulse, like the one shown in Figure \ref{fig:1_Prep_dynamics} (a.i), consists of three RF pulses. After a tip-down pulse of duration $T_{td}$, the magnetization in the simple rotating frame ($\hat{x},\hat{y},\hat{z}$) is

\begin{equation}
    \label{eq:1}
    \textbf{M}(T_{td})=\textbf{R}_{\hat{x}}(-\alpha)\textbf{M}(0). 
\end{equation}

\noindent Here $\textbf{M}(0)=[0,0,M_{0z}]$ is the initial magnetization, and $\textbf{R}_{\hat{x}}(-\alpha)$ represents a matrix rotation around $\hat{x}$ as defined in Supplementary information 1, with $\alpha$ the rotation angle of ideally 90°. Imperfections in the transmitted $B_1$ field can vary significantly across the sample, deviating $\alpha$ from the ideal value depending on the location $\textbf{r}$ by $\alpha(\textbf{r})=\gamma B_1(\textbf{r}) T_{td}$. The second RF applied along $\hat{y}$ is the SL pulse $B_{SL}=\omega_{SL}/\gamma$, of duration $T_{SL}$ with $\omega_{SL}=2\pi F_{SL}$ the induced angular frequency. $\textbf{B}_{SL}$ locks the magnetization in the transverse plane, represented by

\begin{equation}
    \label{eq:2}
    \textbf{M}(T_{td}+T_{SL})=\textbf{R}_{\hat{y}}(-\theta)\textbf{M}(T_{td}),  
\end{equation}

\noindent where $\theta=\omega_{SL} T_{SL}$. In the presence of $B_0$ inhomogeneities, the magnetization will rotate around the effective field $\textbf{B}_{eff}=\frac{\omega_{SL}}{\gamma}\hat{y}+\frac{\Delta\omega_0}{\gamma}\hat{z}$ that forms an angle $\beta = \mathrm{arctan}(\frac{\Delta\omega_0}{\omega_{SL}})$ to $\hat{y}$, where $\Delta\omega_0=\omega_0-\omega_{RF}$ is the off-resonant frequency. The effective rotation angle is then given by $\theta_{eff} = \omega_{eff} T_{SL} = \sqrt{\omega_{SL}^2 + \Delta\omega_0^2} T_{SL}$ \citep{Witschey2007,Yuan2012}. This results in

\begin{equation}
    \label{eq:3}
    \textbf{M}(T_{td}+T_{SL})=\textbf{R}_{\hat{x}}(\beta)\textbf{R}_{\hat{y}}(-\theta_{eff})\textbf{R}_{\hat{x}}(-\beta)\textbf{M}(T_{td}).
\end{equation}
\newline
\noindent Finally, a tip-up pulse of duration $T_{td}$ tilts the magnetization back to the longitudinal axis, in the same way as equation \ref{eq:1} but for a rotation $\textbf{R}_{\hat{x}}(\alpha)$. After preparation, a gradient spoiler in $\hat{z}$ eliminates the remanent transverse magnetization. The spoiler is followed by a slice selective 90° excitation and an echo-planar imaging (EPI) readout, as shown in Figure \ref{fig:1_Prep_dynamics} (a). Although a fast single-shot readout is preferred for functional imaging, the acquisition could be replaced by another sequence. 

We can now consider the dynamic under the influence of an arbitrary target field $B_{NC}(t)$ along $\hat{z}$. In its simplest form, this field is represented by $\textbf{B}_{NC}(t)=B_{NC}\,sin(\omega_{NC}\,t+\varphi)\hat{z}$, with $B_{NC}$, $\omega_{NC}$, and $\varphi$ the amplitude, frequency, and initial phase of the field, respectively. Considering the magnetization direction and the change in the amplitude of the target field constant during each time-step $dt$, the SL period is represented at each instant $\tau$ by a step-wise rotation of the form

\begin{equation}
    \label{eq:4}
    \textbf{M}(\tau) = \textbf{R}_{\hat{x}}(\beta'(\tau))\textbf{R}_{\hat{y}}(-\omega'_{eff}(\tau)dt)\textbf{R}_{\hat{x}}(-\beta'(\tau))\textbf{M}(\tau-dt).     
\end{equation}
\newline
\noindent Where $\beta'$ and $\omega'_{eff}$ are the redefinition of $\beta$ and $\omega_{eff}$ as

\begin{equation}
    \label{eq:5}
    \beta'(\tau) = \mathrm{arctan}(\frac{\Delta\omega_0 + \gamma B_{NC}(\tau)}{\omega_{SL}}),     
\end{equation}
\newline
\noindent and 

\begin{equation}
    \label{eq:6}
    \omega'_{eff}(\tau) = \sqrt{\omega_{SL}^2 + (\Delta\omega_0 + \gamma B_{NC}(\tau))^2}.
\end{equation}
\newline
\noindent If the field is in resonance with the induced SL frequency i.e., $\omega_{NC}=\omega_{SL}$, $\textbf{B}_{NC}$ acts as an excitation pulse and tilts the magnetization away from the SL direction. After the preparation, the drop in the longitudinal magnetization depends on the torque generated by the oscillating target field. We can further generalize Eq. (\ref{eq:4}) to consider that the target field could be along an arbitrary axis (see Supplementary information 1). 

Finally, the magnetization in the presence of relaxation is modified as $\textbf{M}(\tau)=\textbf{A}\,\textbf{M}(\tau-dt)+\textbf{B}$, where the matrices $\textbf{A}$ and $\textbf{B}$ represent the effect of the standard $T_1$ and $T_2$ relaxation times (see Supplementary information 1). In our implementation, each rotation must be infinitesimal and within each step $dt$, two relaxations of $dt/2$ are considered before and after the infinitesimal rotation. In this way, equation \ref{eq:1} becomes:

\begin{equation}
    \label{eq:7}
    \textbf{M}(T_{td}) = \prod_{n=1}^{T_{td}/dt} \textbf{A}[\textbf{R}_{\hat{x}}(-\gamma B_1 dt)\{\textbf{A} \textbf{M}(dt(n-1))+\textbf{B}\}]+\textbf{B},
\end{equation}
\newline
\noindent and equation \ref{eq:4} changes to

\begin{equation}
    \label{eq:8}
    \textbf{M}(T_{td}+T_{SL}) = \prod_{n=1}^{T_{SL}/dt} \textbf{A}_\rho[\textbf{R}_{\hat{x}}(\beta'(\tau))\textbf{R}_{\hat{y}}(-\omega'_{eff}(\tau)dt)\textbf{R}_{\hat{x}}(-\beta'(\tau))\{\textbf{A}_\rho \textbf{M}(T_{td} + dt(n-1))+\textbf{B}_\rho\}]+\textbf{B}_\rho,
\end{equation}
\newline
\noindent where $\textbf{A}_\rho$ and $\textbf{B}_\rho$ represent the relaxations in the rotating frame of reference (see Supplementary information 1). We validated the rotation matrices simulator using the comparison with both analytical and numerical models (see Supplementary information 1).

The dynamic response of the magnetization under the BASL preparation in the presence of a resonant oscillating field is shown in Figure \ref{fig:1_Prep_dynamics} (c.i). 

The RESL preparation divides the SL pulse into two parts with equal duration and opposite phase, as shown in Figure \ref{fig:1_Prep_dynamics} (b.ii). This pulse attempts to cancel out the error in the alpha angle generated by $B_1$ inhomogeneity \citep{Charagundla2003}. Figure \ref{fig:1_Prep_dynamics} (c.ii) illustrates the magnetization dynamics for RESL under the influence of a resonant target field.To simulate this preparation, the sign of $\omega'_{eff}(\tau)$ is inverted in equation \ref{eq:8}, when $n=T_{SL}/2dt$. The sign change adds a phase dependence to the RESL contrast. Depending on the initial phase, and the relation between $T_{SL}$ and $F_{SL}$, the distance to the SL axis gained during the first part of the pulse, could be compensated during the second half. We show this effect in more detail in the following sections.

Finally, the CRESL preparation adds a 180° RF pulse along $\hat{y}$  between the two halves of the SL pulse, i.e., when $n=T_{SL}/2$ and the last pulse is equal to the tip-down. The 180° pulse compensates for the phase accumulated due to the influence of $B_0$ inhomogeneities. Figure \ref{fig:1_Prep_dynamics} (a.iii) shows the pulse sequence, and Figure \ref{fig:1_Prep_dynamics} (c.iii) shows the dynamics of the double resonance effect under this preparation. As in RESL, the CRESL contrast is also dependent on the initial phase, however, the response differs due to the delay in time generated by the 180° pulse.

The double rotating system is generally used to show the magnetization dynamics during an SL pulse \citep{Ito2020,Sogabe2020}. We presented this dynamic in Supplementary information 2.

\subsection{Simulator implementation}

The simulator was programmed in MATLAB (R2019b) using the method presented in section 2.1, together with the rotation and relaxation matrices shown in Supplementary information 1. Parameters that must be given to the simulator related to the sample are the relaxation times of the tissue and the target field as a function of time. Sequence control parameters are $F_{SL}$, $T_{SL}$, $T_R$, $T_E$, number of slices (always 1 for this work) and repetition number. In this way, the entire sequence time can be simulated, considering relaxation in preparation, readout and waiting times (time between slices, preparations without SL).

\subsection{Experimental setup}

We performed measurements in a 3T whole-body MR scanner (Siemens Medical Solutions, Erlangen, Germany), using a Single-Shot-EPI readout with fixed parameters: $T_E=29.8$ ms, Matrix size = 64x64, FOV = 210x210 mm, Bandwidth = 1950 Hz/pixel, and slice thickness = 5 mm. The value of $T_R$ will depend on the selected value of $T_{SL}$. For the simulations, the relaxation times match the values measured on the phantom using commercial sequences. $T_1=1270$ ms and $T_2=200$ ms. We estimated $T_{1\rho}$ using a custom-made sequence that varies the duration of the spin-lock time $T_{SL}$ for a CRESL preparation and fitting the exponential formula $M=M_0 e^{-T_{SL}/T_{1\rho}}$ obtaining $T_{1\rho}=165$ ms.

 \begin{figure}[H]
    \centering
    \includegraphics[width=.8\columnwidth]{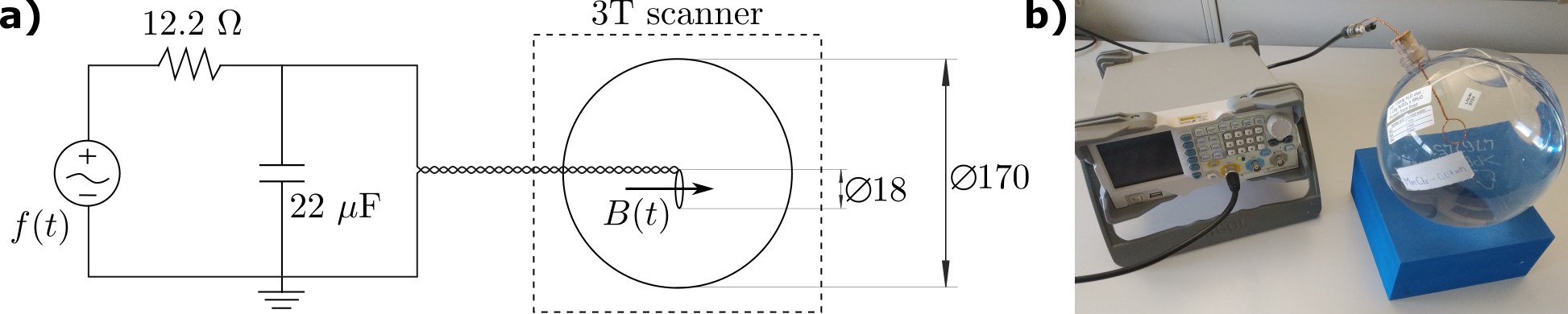} 
    \caption{Diagram of the experimental setup. a) A function generator is connected to a 12.2 Ω resistor in parallel with a 22 nF capacitor to avoid high-frequency coupling. The circuit is connected to a loop coil of isolated copper wire inside a plastic sphere filled with a solution of water and $\mathrm{MnCl}_2$ (0.07 mmol). Diameter measures are in mm. b) Picture of the phantom connected to the function generator outside the scanner room.}
    \label{fig:2_Phantom}
\end{figure}

Figure \ref{fig:2_Phantom} shows the experimental setup used in this study. The phantom consists of a copper loop coil in the middle of a 17 cm diameter plastic sphere filled with saline solution to control the relaxation times. We positioned the phantom such that the loop is in the $xy$ plane and the target oscillating field results in the $B_0$ field direction.

The coil connects to the function generator (Rigol DG1022Z) through the circuit shown in Figure \ref{fig:2_Phantom} (a). The generator allows controlling the amplitude, frequency, and initial phase of the output signal. The external trigger of the scanner gives the signal to the generator to start the output. For the experiments with composite signals, we designed them using MATLAB and imported the data files into the generator using a 1024Sa/s firing rate. 

RESL and CRESL compensate for the influence that  $B_1$ inhomogeneity has on the excitation angle $\alpha$. However, the error in the amplitude of the SL pulse is not corrected and this will affect the  $F_{SL}$ value. To correct for this effect, we performed a manual calibration of the SL amplitude using the CRESL preparation and applied the fitted correction for each SL pulse amplitude used in this work (more detail on Supplementary information 3). 

\subsection{Analysis of SL preparations}

The first set of experiments was designed to compare the performance of the three Spin-lock preparations with respect to initial phase dependence, robustness to field imperfections, and frequency selectivity.

To study the dependence with the initial phase $\varphi$, we simulated the sequence in the resonance condition ($\omega_{NC}=\omega_{SL}=2\pi90$ Hz) as a function of $T_{SL}$ and $\varphi$, for the three preparations shown in Figure \ref{fig:1_Prep_dynamics} (b). For each value of $T_{SL}$ moving between 70 and 100 ms in 1 ms steps, $\varphi$ varies between [-$\pi$,$\pi$] in 0.1 rad steps. The simulations did not consider $B_0$ and $B_1$ inhomogeneities for this analysis.

To better visualize the phase dependency, we simulated and measured the signal contrast as a function of $T_{SL}$ for random and non-random phases. The presented signal contrast was calculated as the ratio between the signal with and without the oscillating initial field ($M/M_0$). For random phases, 100 repetitions were taken and averaged for each $T_{SL}$. For non-random phases, we set the initial phase to 0 and 30 repetitions were acquired and averaged for each point.

To evaluate the robustness of the contrast against $B_0$ and $B_1$ imperfections, we calculated the standard deviation of the contrast for all the possible values of the initial phase $\varphi$ varying between [-$\pi$,$\pi$] in 0.1 rad steps. The map consists of the standard deviation for each combination of the field inhomogeneities. We considered $B_1$ imperfections to deviate the $\alpha$ angle ±5° from the desired 90° value (equivalent to a 0.1 ppm variation around the optimal 100 Hz frequency for a 2.5ms excitation pulse), and $B_0$  inhomogeneity to vary the off-resonance frequency in a [-30,30] Hz range. 

To evaluate the capability of the SIRS preparations to filter specific frequency components, we simulated and measured the contrast for a fixed $F_{SL}$=90 Hz while varying the frequency of the target field. $\omega_{NC}/2\pi$ changed between 60 and 120 Hz in 5 Hz steps with amplitude 75 nT and initial random phase. 100 repetitions were acquired and averaged for each point. We repeated this procedure for each $T_{SL}$ in the range of 70 to 100 ms in 5 ms steps. For each value of $T_{SL}$, we estimated the full-width half maximum (FWHM) and used it as a measure of selectivity of the filter. The simulation was performed for the same parameters, raffling a random phase for each repetition.

\subsection{CRESL in the presence of composite signals}

In this section, we investigated the capability of CRESL to be used as a filter of frequency components when in the presence of composite signals. First, we measured a signal represented by $S(t)=a(t)sin(2\pi53t)+(1-a(t))sin(2\pi97t)$ where the piece-wise constant parameter $a(t)$ changes its value every 30 s intervals assuming the values (0,1/4,1/2,3/4,1). Then, the amplitudes of the 97 and 53 Hz components vary every 30 s but the maximum signal amplitude remains constant throughout the experiment. We chose these frequencies because their separation is greater than the bandwidth of the preparation and because their period is not a multiple of $T_R$, and therefore there is no repeated phases of the oscillation throughout the experiment. The signal was acquired for $F_{SL}$=53 Hz and $F_{SL}$=97 Hz, with $T_{SL}$=100 ms, and $T_R$ = 1017 ms, 206 repetitions were performed that account to 210 s. Measurements consist of successive repetitions alternating $SL_{on}$ and $SL_{off}$ acquisitions. $SL_{on}$ acquisitions were described in section 2.1 and $SL_{off}$ acquisitions are a non-prepared EPI readout where the first $T_{SL}$ ms are left empty. The double resonance effect cannot be separated from other factors influencing the contrast (BOLD, image artifacts, physiological noise, etc.). Here, we propose to use the division of the successive $SL_{on}$ and $SL_{off}$ acquisitions to filter out the spurious effects encoded by the common EPI readout. Therefore, after the acquisition, the point-to-point division ($ppd(t)$) of the signals $SL_{on}$ and $SL_{off}$ was calculated. Each frequency component is computed as the standard deviation of the ppd for each time block.

Delineating the seizure onset zone in refractory epilepsy patients is, in our view, the most relevant clinical application for this sequence. This application requires locating that part of the brain having the highest component of frequencies associated with the epileptogenic activity. Therefore, the second experiment tested the sequence capability to reconstruct frequency components of individually acquired signals. Four signals intend to represent four different voxels of the brain. They all consist of a 60s 0.1 Hz sinusoidal with an amplitude of $x=$81 nT and a 30 Hz sine of amplitude $x/2$ is added the last 30s. For signals 2, 3, and 4, a 90 Hz sine is added with amplitude 0.1x and divided into blocks of 20, 10, and 2 seconds, respectively. To calculate the components of the original signal, we applied a bandpass filter of 20Hz around the target frequency and computed the standard deviation. We measured each signal in time alternating $SL_{on}$ and $SL_{off}$ acquisitions, setting $F_{SL}$ to 30,60 and 90 Hz, with $T_{SL}$=100 ms and $T_R$ = 324 ms. 186 repetitions were taken to account for the 60 s signal duration. We chose these frequencies to comprehend the influence that slow variations of the signal can have on the detection of small amplitude fast oscillations. Finally, we reconstructed the frequency components for each $F_{SL}$ as the standard deviation of the ppd signal.

\section{Results}

\subsection{Analysis of SL preparations}

 \begin{figure}[ht]
    \centering
    \includegraphics[width=.75\columnwidth]{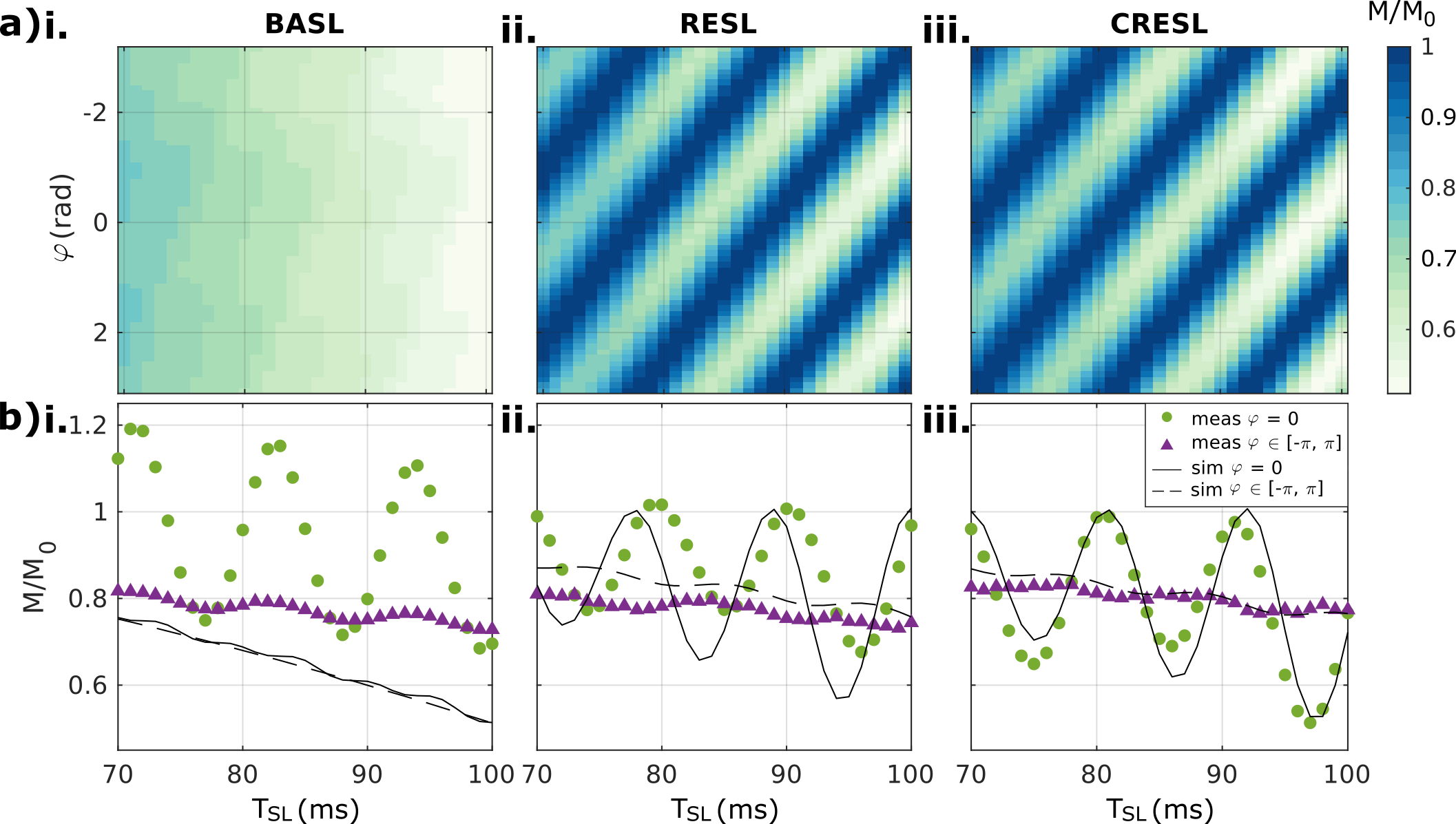} 
    \caption{SIRS contrast dependence with phase. a) Simulated contrast amplitude as a function of  $T_{SL}$ and the initial phase of the target oscillating field $\varphi$ for a sinusoidal field in resonance with the SL pulse for i. BASL, ii. RESL and iii. CRESL. b) Measurement and simulation of the contrast in resonance condition as a function of  $T_{SL}$ for the cases of triggered ($\varphi$ = 0, circles and plain lines) and averaged non-triggered signal (random initial phase, triangles, and dashed lines). $B_{NC}$ = 75 nT, $\omega_{NC}=\omega_{SL}= 2\pi90$ Hz.}
    \label{fig:3_DM_TSL}
\end{figure}

Figure \ref{fig:3_DM_TSL} (a) shows the simulated contrast for the three preparations as a function of $T_{SL}$ and the initial phase of the target field $\varphi$. As seen in Figure \ref{fig:3_DM_TSL} (a.i), BASL is independent of the initial phase and sets the contrast upper limit when all the phases are averaged for a given $T_{SL}$. Figure \ref{fig:3_DM_TSL} (a.ii,iii) demonstrate the phase dependence for RESL and CRESL. Depending on the $T_{SL}$ value, the opposite phase in the second SL pulse refocuses the accumulated contrast, creating the shown stripped pattern. The pattern is shifted between RESL and CRESL due to the delay in $M/M_0$ generated by the 180° pulse. When average contrast in all the phases for RESL and CRESL is half of that of BASL for a given $T_{SL}$. 

Figure \ref{fig:3_DM_TSL} (b) compares the simulated signal without inhomogeneity with measurements for both triggered ($\varphi$=0) and averaged non-triggered initial phase ($\varphi$ varying in the range [-$\pi$,$\pi$]). As shown in Figure \ref{fig:3_DM_TSL} (b.i), field imperfections induce oscillations in BASL, decreasing its average contrast. For certain values of $T_{SL}$, the signal in resonance ($M$) is higher than the signal without resonant field ($M_0$), generating a contrast greater than one. This counterintuitive result can be explained by analyzing the magnetization dynamics for different field inhomogeneity values (a detailed analysis can be found in Supplementary information 4). $B_0$ inhomogeneities also influence the RESL contrast, but the theoretical curve qualitatively represents the experimental signal. For CRESL, instead, the signal corresponds best to its theoretical value for both the phase zero and the mean contrast cases.

\begin{figure}[H]
    \centering
    \includegraphics[width=.75\columnwidth]{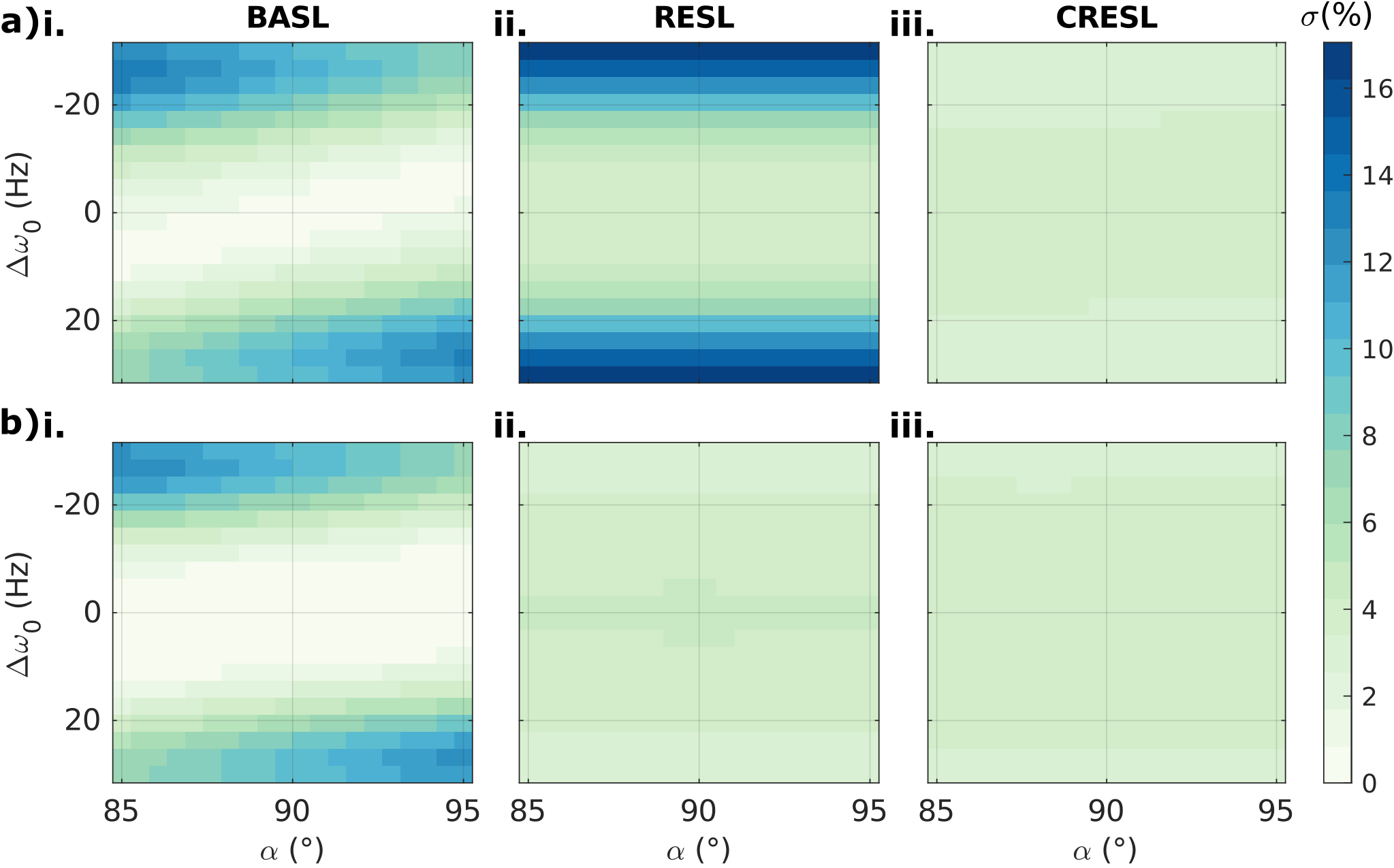} 
    \caption{Simulated contrast variation map (signal standard deviation $\sigma$) as a function of the excitation angle $\alpha$ (°) and the frequency offset $\Delta\omega_0$ (Hz) in resonance condition, $\omega_{NC}=\omega_{SL}= 2\pi90$ Hz. a) For $T_{SL}$=90 ms: i. BASL ($\sigma<13\%$), ii. RESL($\sigma<18\%$), and iii. CRESL($\sigma<4\%$). b) For $T_{SL}$=100 ms: i. BASL($\sigma<13\%$), ii. RESL($\sigma<4\%$), and iii. CRESL($\sigma<4\%$).}
    \label{fig:4_std}
\end{figure}

Given the oscillating behavior showed in Figure \ref{fig:3_DM_TSL}, we defined the signal contrast as the standard deviation of all the random phases. Figure \ref{fig:4_std} shows the variation of the contrast for the three preparations as a function of both $B_0$ and $B_1$ inhomogeneities for two different values of $T_{SL}$ when fixing $F_{SL}$=90 Hz. The maximum std of the contrast is minimum for the CRESL case ($\sigma<4.2\%$ for both cases). RESL is highly dependent on $B_0$ changes and the relationship $T_{SL}$/$F_{SL}$ ($\sigma<17.5\%$ for the worst case). BASL is susceptible to both $B_0$ and $B_1$ inhomogeneities and therefore diagonal patterns are observed ($\sigma<13.4\%$ for both cases). The uncertainty of flip angle between 85 and 95 degree corresponds to that of spin-lock frequency between 85 and 95 Hz. When also considering the influence of the spin-lock frequency error, the maximum variation values are 13.1$\%$, 18.3$\%$ and 4.2$\%$ for BASL, RESL and CRESL respectively. Therefore, this effect is small enough to ignore. This means that CRESL is the most robust against $B_0$ and $B_1$ inhomogeneities when using the standard deviation as the featured metric.

\begin{figure}[H]
    \centering
    \includegraphics[width=.4\columnwidth]{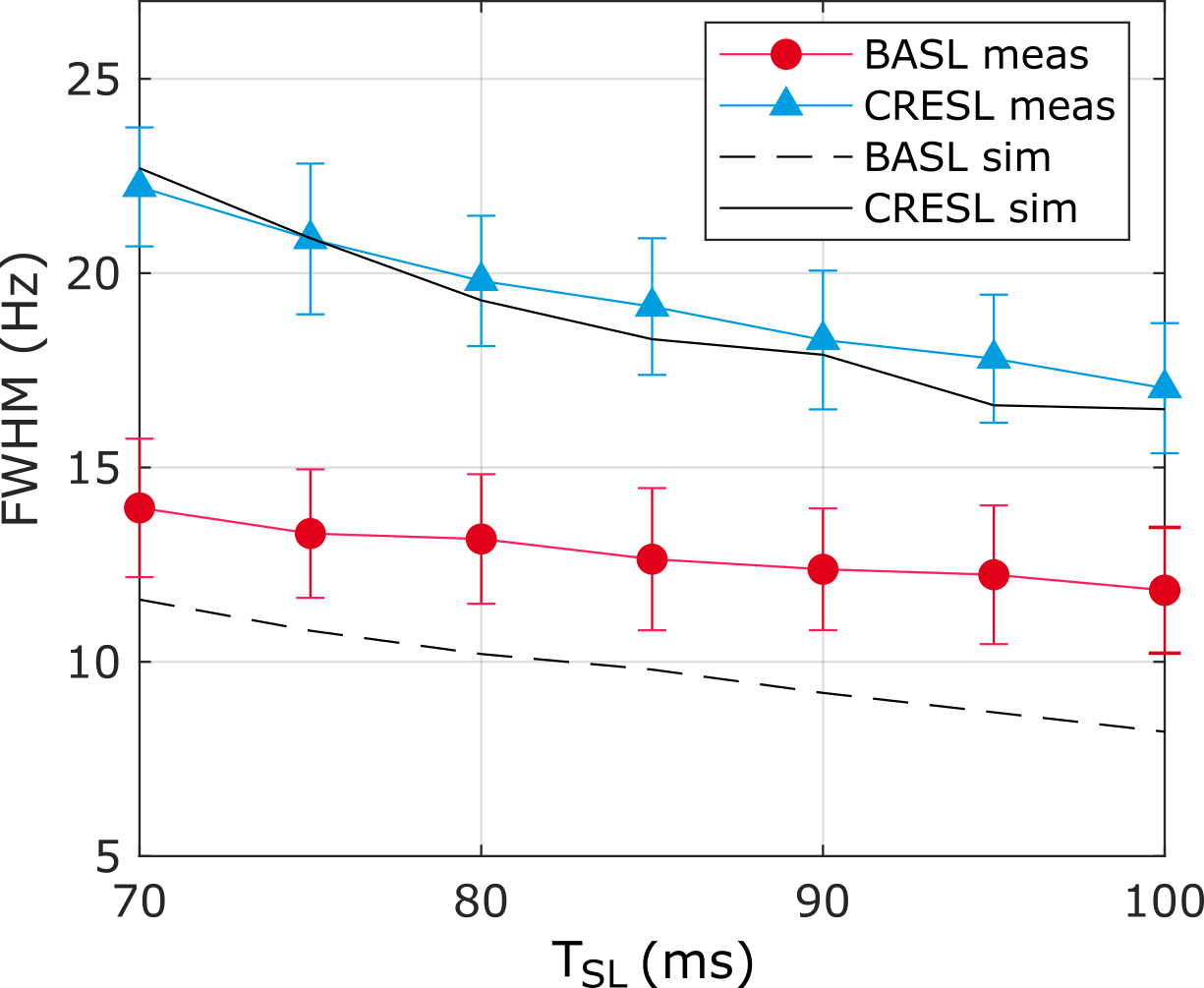} 
    \caption{Simulation and measurement of the FWHM of the double resonance effect for BASL and CRESL preparations as a function of $T_{SL}$. For both cases $B_{NC}$ = 75 nT, $\omega_{SL}= 2\pi90$ Hz.}
    \label{fig:5_BW_TSL}
\end{figure}

Figure \ref{fig:5_BW_TSL} shows the simulation and measurement of the FWHM as a function of $T_{SL}$ for CRESL and BASL. For BASL, the simulation without considering inhomogeneities sub-estimates the measured bandwidth. Still, it remains lower than the CRESL case for the entire $T_{SL}$ range by about 10 Hz.  The FWHM diminishes when increasing $T_{SL}$ for both preparations, i.e., the longer the preparation time, the higher the frequency selectivity. Due to hardware constraints, $T_{SL}$=100 ms is the maximum allowed value, setting the bandwidth to 12 Hz and 17 Hz for BASL and CRESL, respectively. For this $T_{SL}$, and with the minimum possible $T_R$ value, the specific absorption rate (SAR) limits the highest usable $F_{SL}$ to 220Hz (for more details in SAR estimation see Supplementary information 5).

\subsection{Measurements of composite signals}

\begin{figure}[H]
    \centering
    \includegraphics[width=.43\columnwidth]{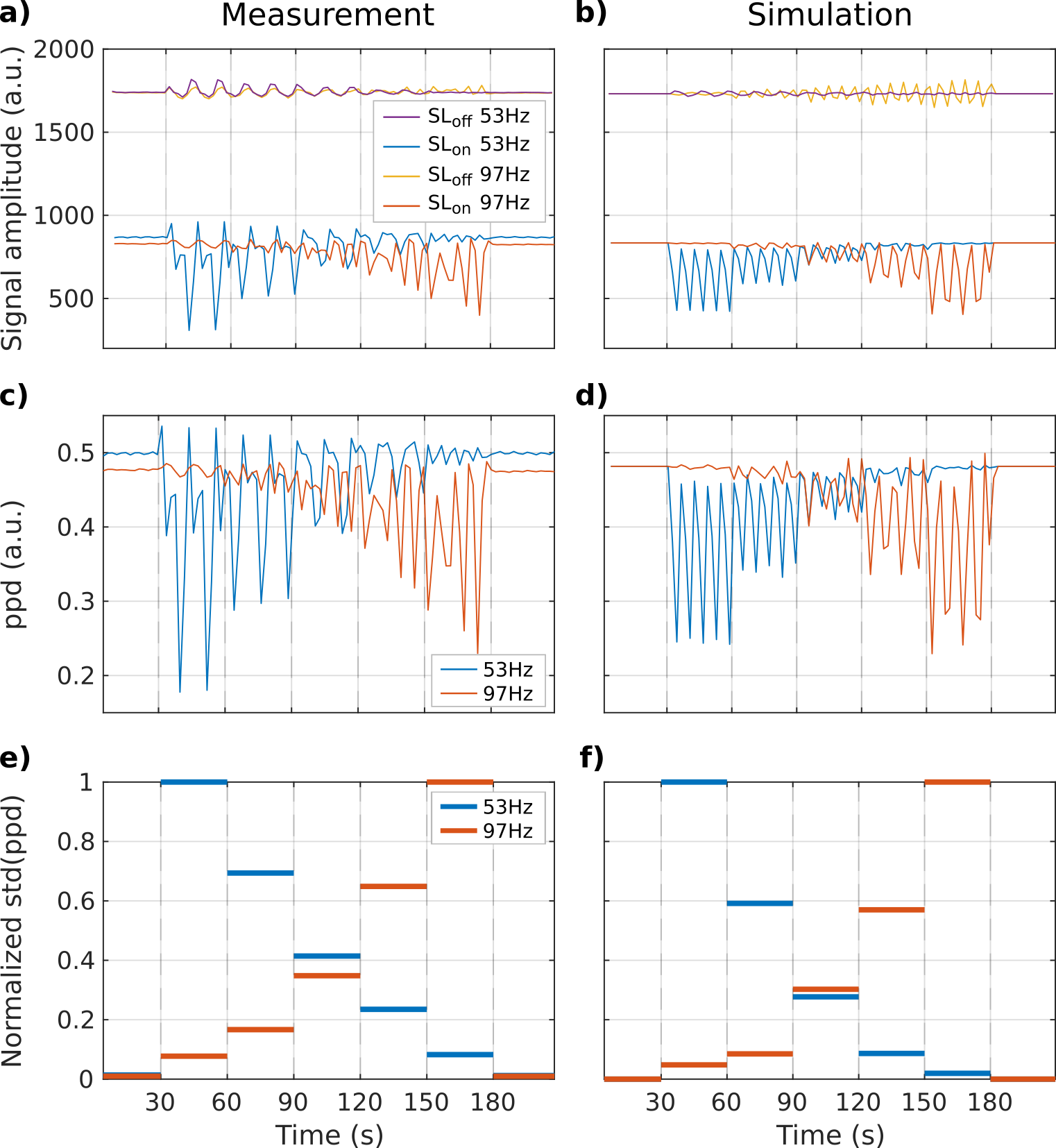} 
    \caption{Sequence response to a composite signal. a) Measured and b) Simulated  $SL_{on}$ and $SL_{off}$ signals as a function of time for $F_{SL}$ set at 53 Hz and 97 Hz. c) Measured and d) Simulated ppd between $SL_{on}$ and $SL_{off}$ as a function of time for the two frequencies used in a). e) Measured and f) Simulated normalized frequency component every 30 s block.}
    \label{fig:6_53_97}
\end{figure}

Figure \ref{fig:6_53_97} (a) shows the measurement of the $SL_{on}$ and $SL_{off}$ signals in time for $F_{SL}$ set to 53 and 97 Hz. The signal behavior is qualitatively well represented by the simulation shown in Figure \ref{fig:6_53_97} (b). Figure \ref{fig:6_53_97} (c) shows the point-to-point division (ppd) between the signals on and off, and Figure \ref{fig:6_53_97} (d) the corresponding simulation. Finally, Figures \ref{fig:6_53_97} (e) and (f) present the measured and simulated normalized components of the two target frequencies. The average error for the normalized std(ppd) between the simulated and measured signal is of 8.6$\%$. This value does not consider the intervals that were normalized to 1. In addition, attempts to fit the curve with the simulator using the initial phase and SL frequency as fit parameters failed to converge.

\begin{figure}[H]
    \centering
    \includegraphics[width=.7\columnwidth]{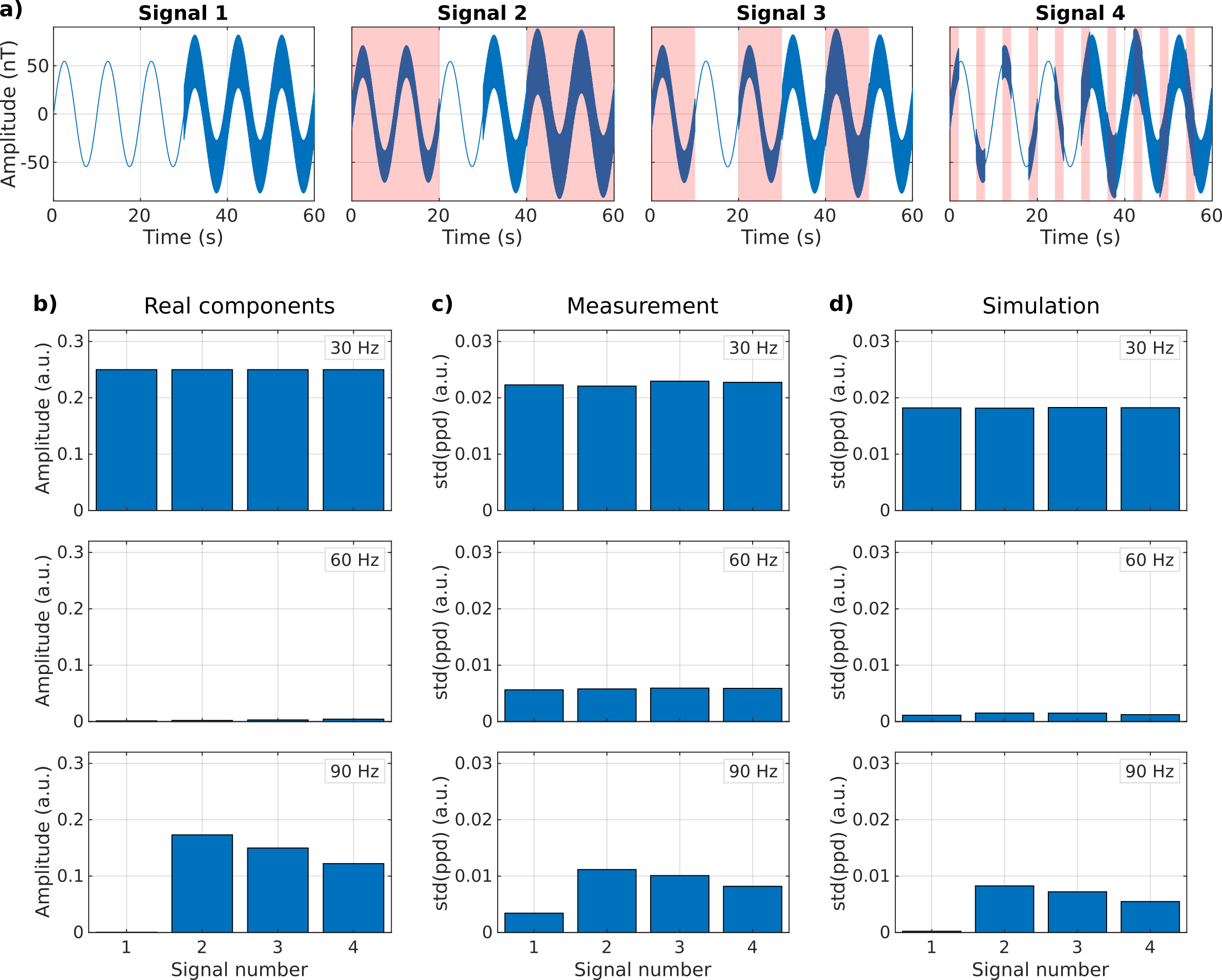} 
    \caption{Component reconstruction of individual signals. a) Input signals composed of a 0.1Hz with a 30 Hz sine the last 30 s. A 90 Hz sine is added in 20, 10, and 2 s blocks for signals 2,3, and 4 respectively, the shaded areas indicate the location of the 90 Hz wave. b) Frequency components of the signals showed in a) at 30, 60 and 90 Hz. c) Components reconstructed from measurement with $F_ {SL}$  fixed at 30, 60 and 90 Hz. d) Frequency components reconstructed from the simulation with the same parameters as in the measurement.}
    \label{fig:7_Artificial_channels}
\end{figure}

Figure \ref{fig:7_Artificial_channels} shows the response of the CRESL sequence to the acquisition of individual signals with different frequency components. Panel a) shows the input signals and panel b) the frequency components corresponding to each signal. Panel c) shows the components reconstructed from the measurement and panel d) from the simulation. The 0.1 Hz ripple appears in the measured signal even if the $F_{SL}$ is 30, 60 or 90 Hz, resulting in a higher ratio in each channel. In the same way, signal amplitude is higher than expected at 60 Hz. From the time-acquired signal (see Supplementary information 6), it is seen that such erroneous detection is caused by traces of the 0.1 Hz oscillation, that decrease for increasing $F_{SL}$. Nevertheless, if the signals represent different volumes in an image, the method can correctly locate the volume with the highest amount of 90 Hz, despite its small amplitude.

\section{Discussion}

We found that the acquisition of the SIRS signal in time will have an oscillating behavior and the reason behind the oscillations depends on the type of SL preparation. According to Figure \ref{fig:3_DM_TSL} (a), the rotary preparations will oscillate due to the phase of the oscillating field and the relation between $T_{SL}$ and $F_{SL}$, as was previously suggested by \citep{Jiang2016}. In addition, the theoretical BASL signal does not depend on the phase and presents the contrast upper limit when averaging in all the possible phases. However, Figure \ref{fig:3_DM_TSL} (b) shows that in practice, $B_0$ and $B_1$ inhomogeneities induce oscillations also for the BASL approach, deviating the contrast from the theoretical value. RESL has a mixed behavior, being dependent on $B_0$ and the initial phase of the oscillating field and finally, CRESL oscillates exclusively due to the initial phase. 

The optimal contrast to locate oscillating fields should depend solely on the double resonance effect and not on the spatial variation of the $B_0$ and $B_1$ fields. Given the unavoidable oscillating nature of the contrast, and as it was previously suggested for the SLOE approach \citep{Jiang2016,Coletti2021}, we adopted the standard deviation as a metric to quantify the influence of the target field. Within this metric, according to Figure \ref{fig:4_std}, CRESL is the most robust preparation even within the double resonance dynamic, which further supports the idea firstly proved in the field of $T_1\rho$ relaxation \citep{Witschey2007,Wang2015}. On the other hand, RESL still depends on the heterogeneities of $B_0$ (Figure \ref{fig:4_std} (a.ii)), and for certain combinations of $T_{SL}$ and $F_{SL}$, RESL presents the highest variability. Therefore, we have discarded it in the following analysis.

The analysis of the bandwidth presented in Figure \ref{fig:5_BW_TSL} shows that BASL is in average more selective than CRESL for about 10 Hz in the studied range. This would indicate that BASL can better separate closer frequency components of a composite signal. In addition, Figure \ref{fig:5_BW_TSL} shows that the bandwidth can be reduced by increasing $T_{SL}$. Nevertheless, the Specific Absorption Rate (SAR) and the scanner's hardware constraints will limit the highest $T_{SL}$ value. To avoid these limitations, $T_R$ must grow, but this means lowering the temporal resolution when signals are acquired in time. Therefore, the trade-off between increasing frequency selectivity and lowering the temporal resolution must be considered for future applications (further SAR analysis is presented in Supplementary information 5).

Based on the results of the analysis of the SL preparation, we decided to adopt the CRESL method due to its robustness. Using this preparation, we tested the response to composite signals by acquiring the signal over time. When analyzing a signal that varies the amplitude of its frequency components in a step-wise manner over time, the simulator and the measured signal showed the same empirical behavior, as is shown in Figure \ref{fig:6_53_97} (a) and (b).  However, there are several differences between them in terms of waveform and the number of peaks and the reconstructed components showed in Figure \ref{fig:6_53_97} (e) and (f) present a variation of 8.6$\%$. Furthermore, a measurement fit using the SL frequency and the initial phase of the field as free parameters failed to converge. Although this analysis is limited to a qualitative evaluation only, both ppd signals showed changes in amplitude that represent the expected changes in the ratios of each block. 

Our last experiment aimed to test the sequence capability to reconstruct frequency components of individually acquired signals without including prior information during post-processing. As shown in Figure \ref{fig:7_Artificial_channels}, the high amplitude of the low frequency variation induces a detection despite not being in resonance (see also Supplementary information 6). This effect was not seen in the simulator. We hypothesize that this effect is because the simulator does not consider the influence of the sequence during image acquisition, which affects both the acquisition with and without SL. This reveals a limitation of the proposed ppd metric. The point-to-point division of the signals on and off is designed to eliminate artifacts detected in the readout part of the sequence, however, those artifacts will only be filtered out when they are either static or vary slower than $2T_R$ (time it takes to acquire two consecutives on and off signals). Despite this, the proportions shown follow the same behavior as the frequency components. The signal manages to reconstruct which channel has the highest amount of high-frequency component. This proof of principle experiment is promising for the application of this sequence to create a map where each pixel amplitude represents the activation at a given frequency. It also shows that it is necessary to make the simulator more complex to include the effect of image acquisition, as has been previously proposed \citep{Ueda2021}.

In summary, there were two main findings in this work. First, the double resonance signal is oscillatory for the three studied preparations. From the analysis of these oscillations, it was found that CRESL is the most robust against field inhomogeneities when using the std(ppd) as a metric. Second, the ppd signal can separate and estimate the amplitude of frequency components within individually acquired composite signals. However, there are two strong limitations of this method. The first one, common to the three SL preparations, is that the susceptibility of the SL amplitude to  $B_1$ is countered with the SL calibration but not corrected. This calibration requires extra measurements. The second one is that multiples acquisitions are necessary for a reliable calculation of the std(ppd). This in turn requires that the resonant oscillating field is present during sufficiently long periods of time. This would require the analysis of more realistic signals to evaluate the clinical usefulness. These results suggest that the metric and preparation studied can be used in future research for the characterization of physiological signals.

\section{Conclusions}

We investigated the magnetization dynamics of three Spin-lock preparations in the presence of a resonant oscillating field and characterized the influence of field inhomogeneities. By simulations and phantom experiments, we found that all three spin-lock preparations exhibit an oscillating contrast as a function of time.  Due to the oscillating behavior, we used the standard deviation of the signal as a metric to quantify the influence of the target field. We showed the std metric within the CRESL preparation is the most robust against field inhomogeneities even in the presence of the double resonance effect. 

By acquiring composite signals with the chosen CRESL preparation, we found that is possible to distinguish the frequency components of individual signals composed of multiple frequencies. These results indicate that the CRESL-prepared SIRS sequence has the potential to be used for the detection and localization of oscillating fields with frequencies associated with specific pathologies. In the future, we aim to assess the detection limit of the presented methods under realistic physiological and pathological conditions.

\bibliographystyle{unsrt}
\bibliography{library}

\section*{Acknowledgements}
This study was supported by the Swiss National Science Foundation (SNSF) within the project: Predict and Monitor Epilepsy After a First Seizure: The Swiss-First Study (SNSF, CRSII5-180365, PI Roland Wiest), the Kommissionspräsident Kernen Fonds for neurological research and the Schweizerische Epilepsie‑Liga (Swiss league against epilepsy).

We thank Hans Slotboom, Guodong Weng and Andreas Federspiel (Institute for Diagnostic and Interventional Neuroradiology, Support Center for Advanced Neuroimaging (SCAN), University of Bern, Bern, Switzerland) for their essential discussion of the work. We also thank Regina Reissmann (Biomedical analyst, NEUR, Inselspital, Bern) and Thierry Rumo (Medical technician, DT, Inselspital, Bern) for the technical help in the implementation of the phantom.

\section*{Author contributions}
All authors conceived the idea of this work. M.C. designed the experiments, implemented the simulations, and wrote the manuscript. M.C. and F.T. performed the measurements and analyzed the data. C.K. and R.W. supervised this study. All authors provided discussion and reviewed the manuscript.

\section*{Data availability statement}
Simulation source codes and data can be shared upon reasonable request to M.C.

\section*{Additional Information}
The authors declare no competing interests.

\section*{Supplementary Information}

\subsection*{1. Simulator details}

The used rotation matrices $\textbf{R}_{\hat{x}}(\alpha)$, $\textbf{R}_{\hat{y}}(\alpha)$, $\textbf{R}_{\hat{z}}(\alpha)$ can be represented by
\begin{equation}
\textbf{R}_{\hat{x}}(\alpha) =  
\begin{bmatrix} 1 & 0 & 0\\ 0 & cos(\alpha) & -sin(\alpha) \\ 0 & sin(\alpha) & cos(\alpha) \end{bmatrix}, \textbf{R}_{\hat{y}}(\alpha) =  
\begin{bmatrix} cos(\alpha) & 0 & sin(\alpha)\\ 0 & 1 & 0 \\ -sin(\alpha) & 0 & cos(\alpha) \end{bmatrix}, \textbf{R}_{\hat{z}}(\alpha) =  
\begin{bmatrix} cos(\alpha) & -sin(\alpha) & 0\\ sin(\alpha) & cos(\alpha) & 0 \\ 0 & 0 & 1 \end{bmatrix}.
\nonumber
\end{equation}

\noindent Each rotation appears counter-clockwise when the axis about which they occur points toward the observer, and the coordinate system is right-handed.

\noindent The relaxation matrices $\textbf{A}$, $\textbf{B}$, $\textbf{A}_\rho$ and $\textbf{B}_\rho$ can be represented by

\begin{equation}
\textbf{A}\!=\!  
\begin{bmatrix} e^{-t/T_2} & 0 & 0\\ 0 & e^{-t/T_2} & 0 \\ 0 & 0 & e^{-t/T_1}\!\end{bmatrix}\!\!,        \textbf{B}\!=\!\begin{bmatrix} 0 \\ 0 \\ 1-e^{-t/T_1} \end{bmatrix}\!,                          \textbf{A}_{\rho}\!=\!\begin{bmatrix} e^{-t/T_{2\rho}} & 0 & 0\\ 0 & e^{-t/T_{1\rho}} & 0 \\ 0 & 0 & e^{-t/T_{2\rho}} \end{bmatrix}\!,
\textbf{B}_{\rho}\!=\!\begin{bmatrix} 0 \\ M_{\infty,\hat{y}}(1-e^{-t/T_{1\rho}}) \\ 0 \end{bmatrix}\!.
\nonumber
\end{equation}

\noindent $M_{\infty,\hat{y}}$ is the equilibrium magnetization in the double rotating frame along the spin-lock axis, which we neglect for being much smaller than the generated by the static field $B_0$. Within the simulator, each rotation step of duration $dt$ will have two relaxation steps of $dt/2$, which means evaluating the four relaxation matrices in $t=dt/2$. 

\noindent To consider that the target field could oscillate along an arbitrary axis, equation (\ref{eq:4}) can be generalized as 

\begin{equation}
    \textbf{M}(\tau) = \textbf{R}_{\hat{z}}(\delta(\tau))\textbf{R}_{\hat{x}}(\zeta(\tau))\textbf{R}_{\hat{y}}(-\omega'_{eff}(\tau)dt)\textbf{R}_{\hat{x}}(-\zeta(\tau))\textbf{R}_{\hat{z}}(-\delta(\tau))\textbf{M}(\tau-dt).     
\nonumber
\end{equation}

\noindent Where $\delta$ and $\zeta$ represent the angles formed by the effective field direction with respect to $\hat{y}$ in the yx and yz plane respectively. It follows that when the direction of the target field is along $\hat{z}$, $\delta$ is 0 and we recover equation (\ref{eq:4}).

To validate the rotation matrix simulator (RM sim), we compared it with the analytical model presented by Ueda et al. 2018 and the numerical solution of the differential equation presented in this same work using the 4th-order Runge-kutta. The analytical simulator is only valid for a sinusoidal field, so we compared the three simulated signals as a function of the amplitude of the target field. As shown in Supplementary Figure 1. a), both numerical methods approximate well to the Analytical solution. However, the Runge-kutta methods needs a smaller time step ($dt <$ 5E-4 s) to match the accuracy of the RM simulator, as can be seen in panel b). For the same time step $dt$, the calculation time of the Runge-kutta method is slightly smaller. However, to match the same percentual error, dt must be lower than the required for the RM simulator, making the effective calculation time bigger. 

 \begin{figure}[H]
    \centering
    \includegraphics[width=.7\columnwidth]{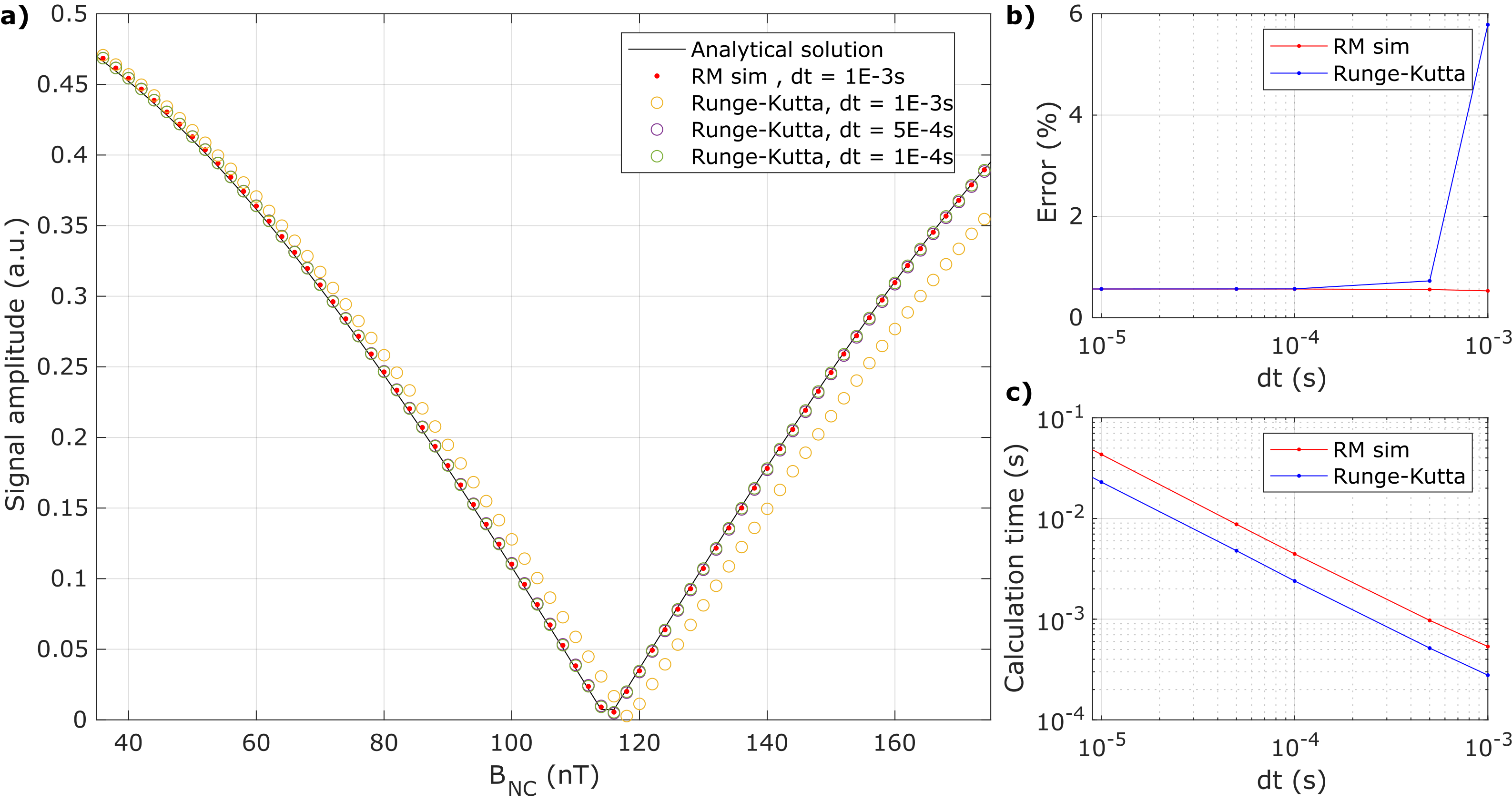} \par
    {\small Supplementary figure 1: Validation of the proposed simulation method. a) Signal as a function of the target field amplitude simulated using the analytical solution to the differential equation, the rotation matrix simulator used in this work, and the numerical solution obtained with a 4th order Runge-kutta method. The Runge-kutta was calculated for different times steps until the error (calculated as the mean distance to the analytical solution) was lower than 1$\%$. b) Percentual error as a function of time step for the rotation matrix simulator and 4th order Runge-kutta solution. c) Calculation time as a function of time-step for the proposed rotation matrix simulator and 4th order Runge-kutta solution.}
\end{figure}

\subsection*{2.	Dynamics in the double rotating frame}

The double rotating frame is usually employed to depict magnetization dynamics under SL preparation techniques. The double rotating frame (x’,y’,z’) is defined as if the simple rotating frame is now also rotating around the SL axis at the SL frequency. We can decompose the target oscillating field $\textbf{B}_{NC}(t) = B_{NC} sin(\omega_{NC} t + \varphi) \hat{z}$ into two fields of half the amplitude oscillating in the xz plane with opposite frequencies as shown in Supplementary Figure 2 i): 

\begin{equation}
     \textbf{B1}_{NC}(t) = \begin{cases} B_{NC}/2.cos(\omega_{NC}t+\varphi)\hat{x} \\ 0 \\ B_{NC}/2.sin(\omega_{NC}t+\varphi)\hat{z} \end{cases} \quad and \quad \textbf{B2}_{NC}(t) = \begin{cases} B_{NC}/2.cos(-\omega_{NC}t-\varphi + \pi)\hat{x} \\ 0 \\ B_{NC}/2.sin(-\omega_{NC}t-\varphi + \pi)\hat{z} \end{cases}
\nonumber
\end{equation}

When in resonance with the SL frequency, one of the fields will be static in the double rotating frame while the other one will be oscillating at $2\omega_{SL}$. Applying the rotating wave approximation (firstly proposed in this context by Ueda et al., Journal of Magnetic Resonance, 2018), we discard the effect that the off-resonance field will have on the magnetization. Supplementary Figure 2 ii), iii) and vi) shows the magnetization dynamics for the three presented SL preparations in the double rotating frame. For BASL there is only one SL frequency, therefore the target field $\textbf{B1}_{NC}(t)$ remains static and deviates the magnetization from the SL axis. For RESL and CRESL, the same happens during the first half of the SL pulse. When the SL changes sign, the opposite field $\textbf{B2}_{NC}(t)$ will be on resonance, and its initial position will depend on the phase accumulated during the first part of the evolution. Depending on this phase, the contrast accumulated during the first SL part can be compensated.

 \begin{figure}[H]
    \centering
    \includegraphics[width=.85\columnwidth]{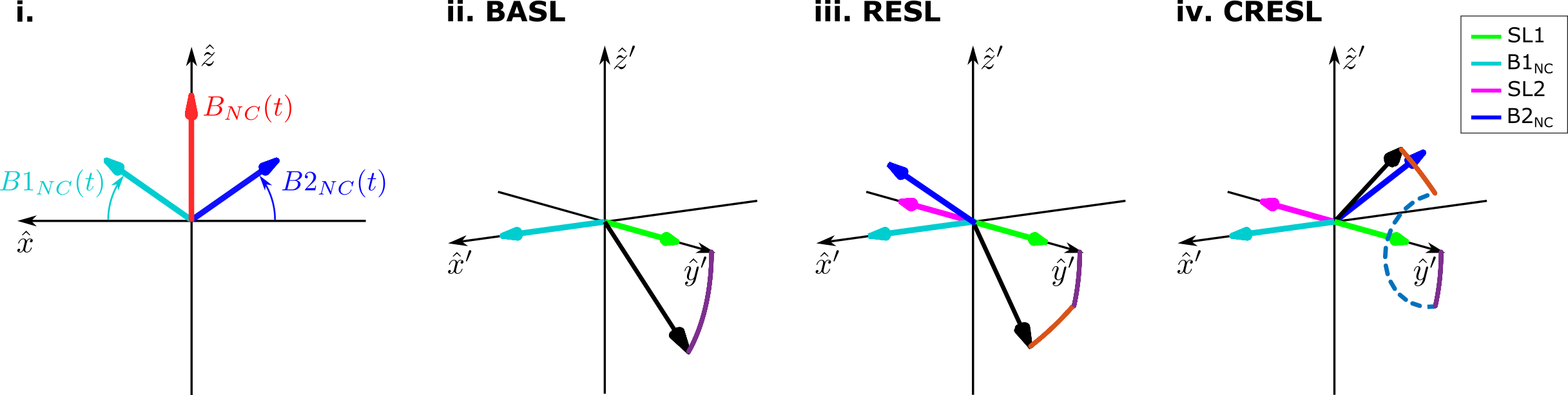} \par
    {\small Supplementary figure 2: Magnetization dynamics in the double rotating frame. i) Target field decomposition in the simple rotating frame. ii), iii) and iv) BASL, RESL and CRESL magnetization dynamics represented in the double rotating frame respectively. As the double rotating frame rotates around the SL axis, the sign of rotation changes when SL1 changes to SL2 in the RESL and CRESL cases. Therefore, only SL1 and $B1_{NC}$ are present during the first SL part, and SL2 and $B2_{NC}$ are present during the second part for RESL and CRESL. The magnetization path follows the same color code as the one used in Figure \ref{fig:1_Prep_dynamics} of the main text. The 180° pulse dynamic is indicated with dashed lines.}
\end{figure}

\subsection*{3. SL pulse calibration}

An error in the excitation angle can be corrected by the change in the sign of the second half of the spin-lock pulse (RESL) and an error in the $B_0$ field can be corrected by the 180° pulse (CRESL). However, no compensation can be made for the $B_1$ inhomogeneity influence during the application of the SL pulse. This will induce a $F_{SL}$  different from the one of the target field and the resonance condition will not be fulfilled. To solve this problem, an SL calibration can be made to find the $B_1$ error for each target frequency. Supplementary Figure 3 a) shows the measurements at different resonant frequencies as a function of $F_{SL}$ and b) the obtained linear regression. The initial phase was set to 0, and 30 repetitions were acquired and averaged for each point on each curve. From the fitted curve, the SL amplitude was corrected before every measurement.
    
 \begin{figure}[H]
    \centering
    \includegraphics[width=.7\columnwidth]{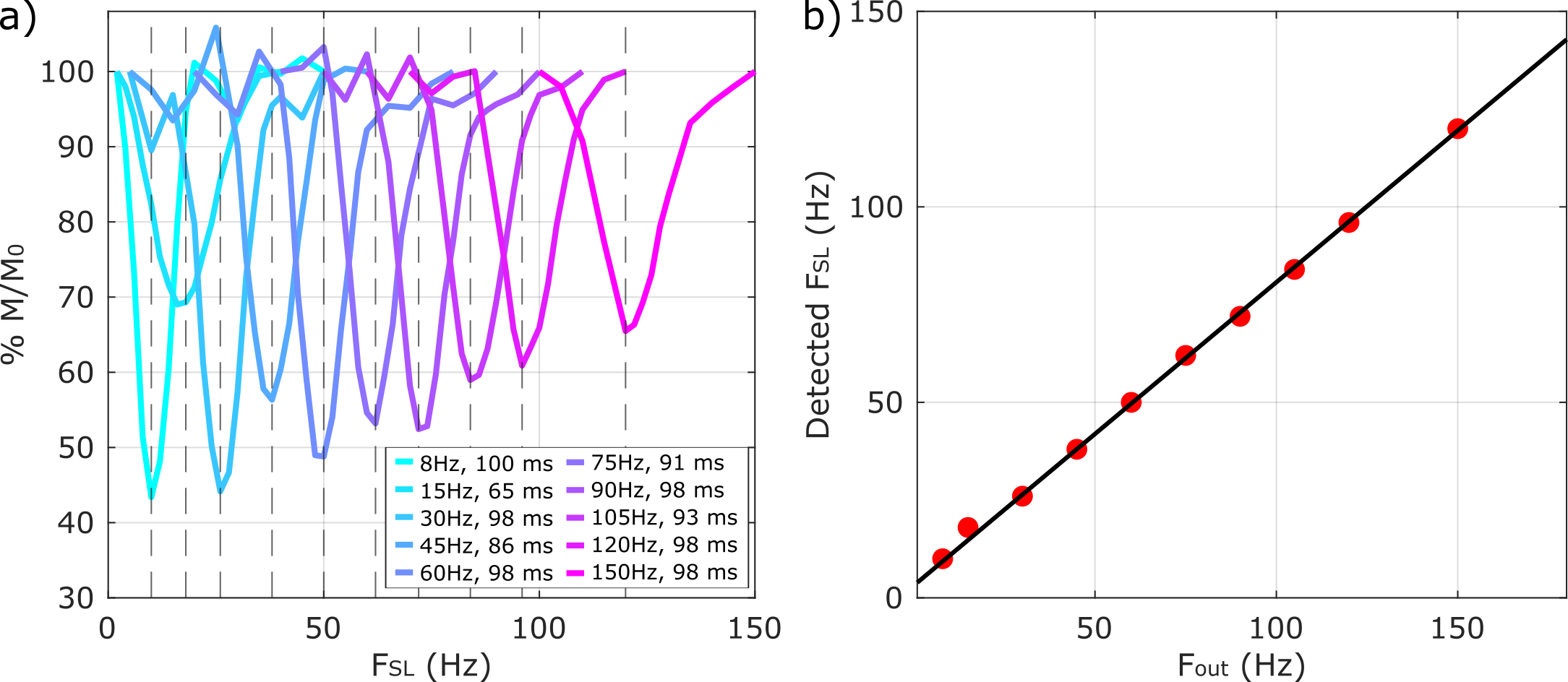} \par
    {\small Supplementary figure 3: SL amplitude calibration. a) Percentual change of the signal as a function of $F_{SL}$ for different values of the target oscillating field frequency ($F_{out}$). b) Linear regression of the detected ($F_{SL}$) vs the expected frequency ($F_{out}$). The obtained slope was then used to define the amplitude of each SL pulse.}
\end{figure}

\subsection*{4.	BASL contrast in the presence of field inhomogeneities}

To explain the result showed in Figure 3 (b.i) of the main text, we fitted the BASL contrast using the field inhomogeneity as fitting parameter. The excitation angle $\alpha$, the off-resonance frequency $\Delta \omega_0$ and the spin locking frequency $F_{SL}$ were used as fitting parameters. The rest of the parameters were set as stated in the methods. Supplementary Figure 4 shows the fitting result. The fitted parameters were $\alpha$=73°, $\Delta \omega_0$=0 Hz, and  $F_{SL}$=90,88 Hz. This indicates that $B_1$ is the main contributor to the behavior observed in this case. This behavior comes from the magnetization without a target field ($M_0$), when the tip down pulse angle is smaller than 90°. This generates an oscillation around the SL axis that result on a transversal magnetization after the tip up pulse.

 \begin{figure}[H]
    \centering
    \includegraphics[width=.5\columnwidth]{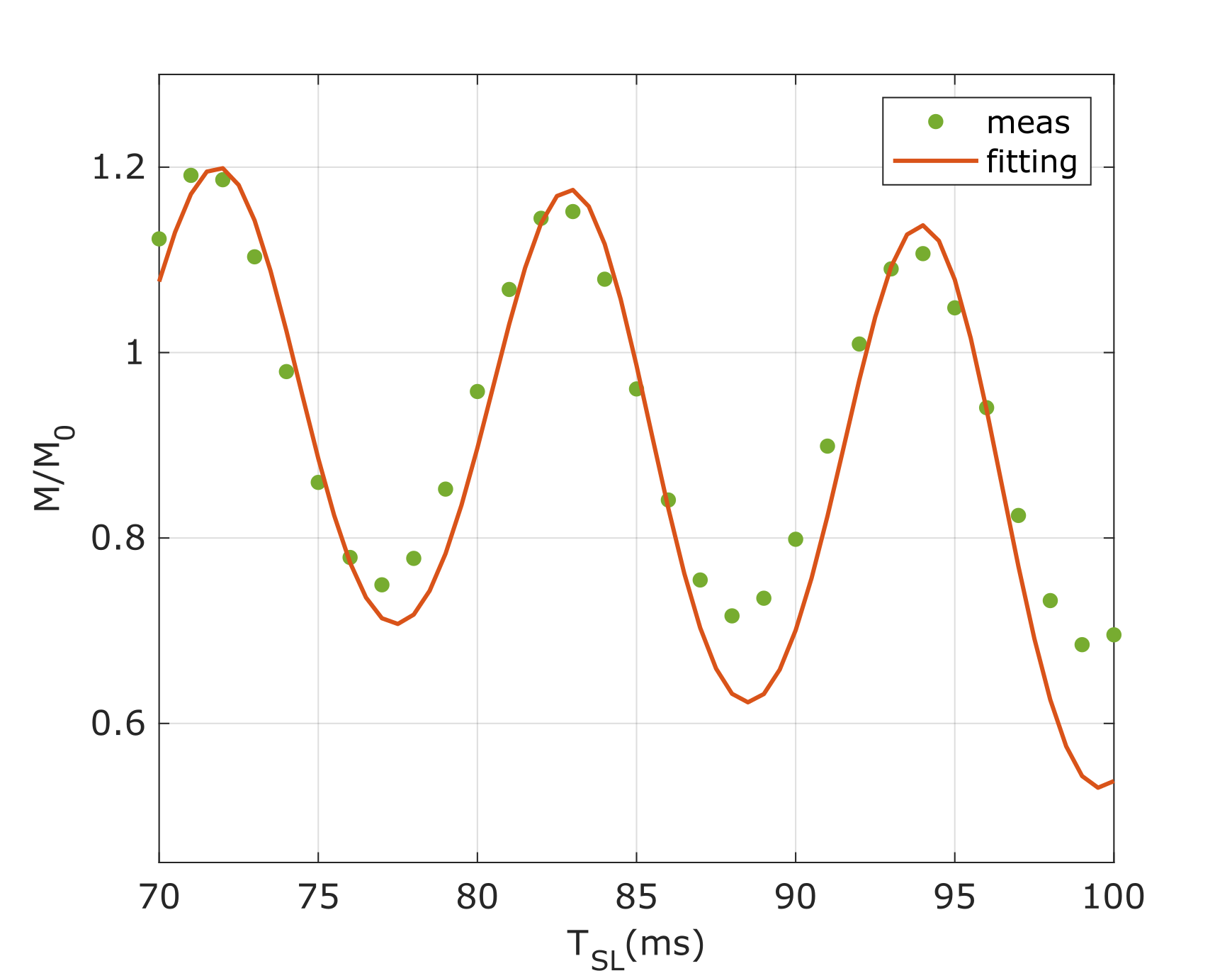} \par
    {\small Supplementary figure 4: Measured BASL contrast as a function of the SL time and fitted curve.}
\end{figure}

\subsection*{5.	SAR estimation}

The specific absorption rate (SAR) limits the range of detectable frequencies for the SIRS approach, especially at high field strengths. Previous reports have calculated the SAR of the CRESL sequence; however, the values are highly dependent of sequence definition, coil used and subject scanned. Therefore, to assess the detection limits for our own experimental set-up, we used the SAR check computed by the scanner. In our case, we used a 64-channel head birdcage coil and set the subject values to typical values of high (170 cm) and weight (60 kg). The minimum $T_R$ is limited by $T_{SL}$. Using the maximum allowed value, $T_{SL}$=100 ms, we would have the highest temporal resolution and worst SAR case for the minimum possible $T_R$=165 ms (minimum possible value given by preparation duration plus duration of the EPI readout). For this value, the obtained SAR as a function of $F_{SL}$ is shown in Supplementary Figure 5.

The SAR value is the same for BASL and RESL, only differing with CRESL by the 180° pulse. The greatest contribution to SAR is given by the SL pulses, therefore it does not differ so much between the proposed preparations. In this setup, the frequency limit before exceeding the limit set by the International Electrotechnical Commission (IEC) (3.2 W/Kg for the head) is 222 Hz, 220 Hz and 310 Hz for BASL and CRESL and the  $SL_{on}$ (CRESL)/$SL_{off}$ approach respectively. If a higher frequency value must be used as target, $T_R$ needs to be increased with the consequent loss of temporal resolution.

 \begin{figure}[H]
    \centering
    \includegraphics[width=.45\columnwidth]{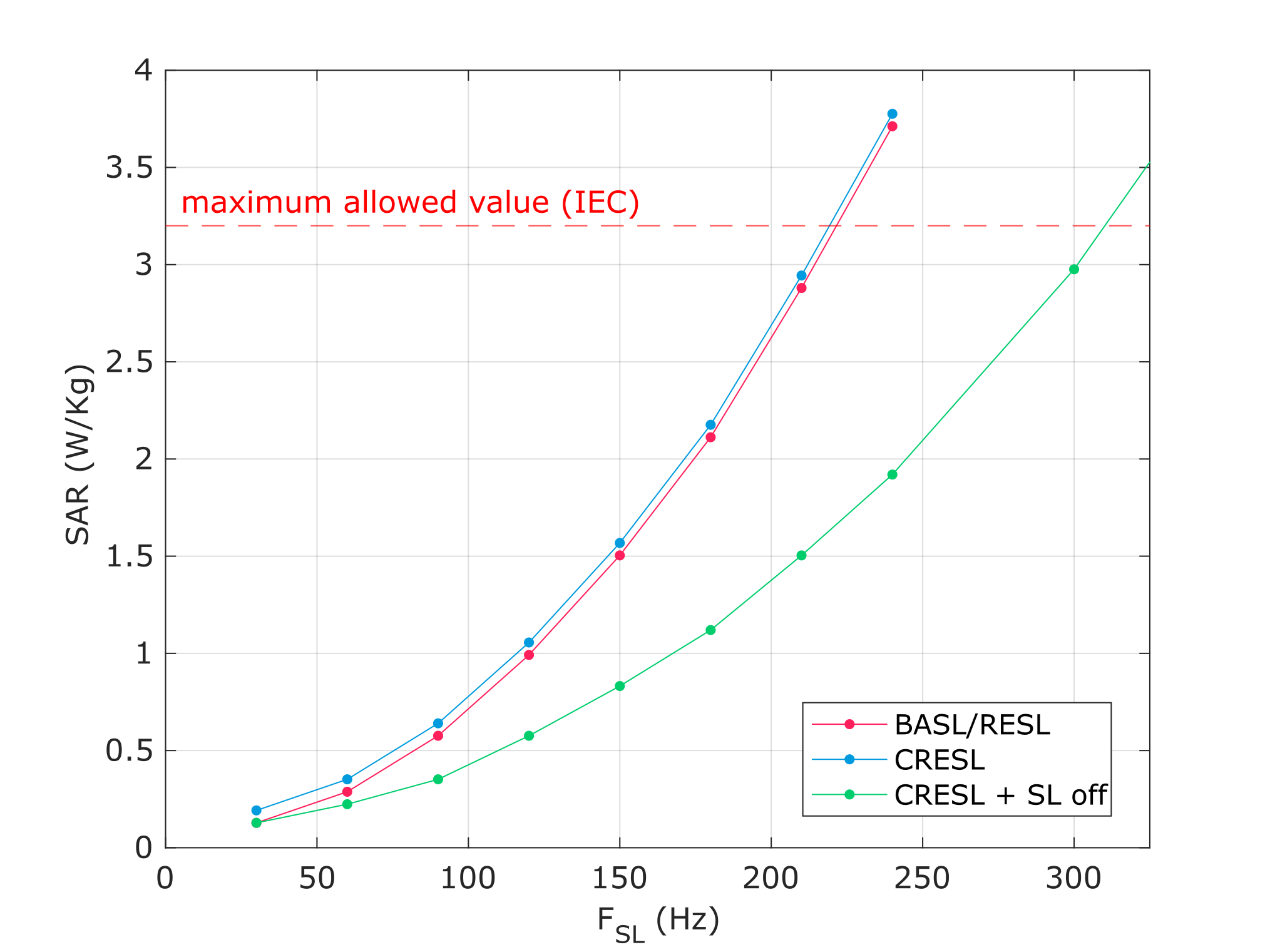} \par
    {\small Supplementary figure 5: SAR estimation. Estimated SAR as a function of $F_{SL}$  for both BASL and RESL, CRESL and the combined $SL_{on}$/$SL_{off}$ approach. For all the sequences, $T_{SL}$=100 ms, $T_R$=165 ms, and the rest of the parameters are the same as described in part 2.3 of the Methods section.}
\end{figure}

\subsection*{6. Measurement of individual signals}

Supplementary Figure 6 shows the point-to-point division between the $SL_{on}$ and $SL_{off}$ acquisition of the 4 composite signals presented in Figure \ref{fig:7_Artificial_channels} a). The signal shows traces of the 0.1 Hz oscillation, despite the acquisition being performed for $F_{SL}$ equal to 30, 60 and 90 Hz. This influence is greater at 30 Hz and diminishes when increasing $F_{SL}$. This shows that slow drifts cannot be fully filtered out using the ppd between on and off acquisitions. The reason behind this is that the two acquisitions have a time difference of TR. Therefore, any signal variation that occurs within a $T_R$ will be registered differently in the $SL_{on}$ and $SL_{off}$ signal. An alternative to this is applying a high-pass filter to eliminate the influence of slow drifts if there is prior knowledge of the signals present. Nevertheless, the 90 Hz signal only has $10\%$ of the amplitude of the slow-wave and is still the biggest detected component at $F_{SL}$=90 Hz.

 \begin{figure}[H]
    \centering
    \includegraphics[width=.7\columnwidth]{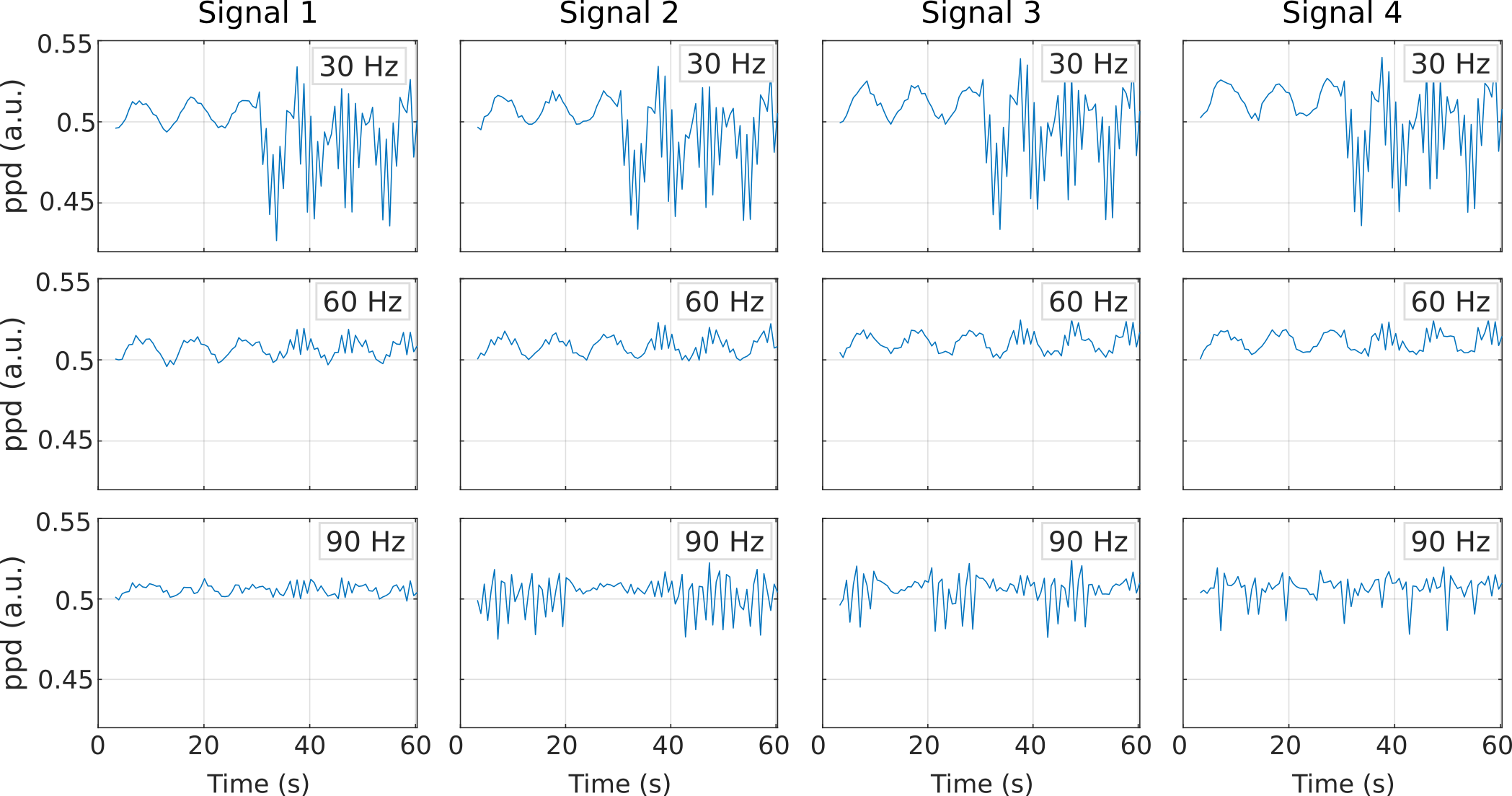} \par
    {\small Supplementary figure 6: Point to point division (ppd) between $SL_{on}$ and $SL_{off}$ measured for the input signals shown in Figure \ref{fig:7_Artificial_channels} a). Each column corresponds to each of the input signals and each row to the ppd acquired at $F_{SL}$ 30, 60 and 90 Hz.}
\end{figure}

\end{document}